\documentclass[preprint,aps,prc,showpacs,preprintnumbers,amsmath,amssymb]{revtex4}

\usepackage{graphicx}
\usepackage{dcolumn}
\usepackage{bm}
\usepackage{ulem}

\begin{document}

\title{Isotopic Yields and Isoscaling in Fission}

\author{W.A. Friedman} 
\affiliation{Department of Physics, University of Wisconsin,
Madison, WI 53706, USA}

\date{\today}

\begin{abstract}

A simple model is proposed to examine the isotopic yields of the
fragments from binary fission.
For a given charge partition the peaks and widths in the isotope distributions
are studied both with the liquid-drop model and with shell modifications.  
The basis for isoscaling is also explored. The  symmetry energy 
plays a dominant role in both the distributions and the isoscaling behavior.
A systematic increase in the isoscaling parameter,
$\alpha$, with the proton number of the fragment element is 
predicted in the context of the liquid-drop model. Deviations
arising from shell corrections are explored.
\end{abstract}

\maketitle
\newpage
    Recent studies have explored the role of the symmetry energy in
governing the isotope yields in a variety of nuclear
reactions,
including deep-inelastic collisions, evaporation following
excitation, and multifragmentation\cite{ts1,
ts2,ts3}. This previous work
examined  the basis
for the experimental signal of isoscaling. 
That signal  is seen when
ratios are calculated for 
the yields of isotopes from reactions in systems of 
similar energy but different N/Z values. The isoscaling signal 
is present when these ratios, $R_{21}(n,z)$, display
the simple form 
\begin{equation}
R_{21}(n,z) \propto {\rm exp}(\alpha n + \beta
z).
\end{equation}
Here $n$ and $z$ are the neutron and proton numbers of each nuclide
produced in the two reactions for total systems which are characterized by 
the the labels 2 (heavier) and 1 (lighter).
Isoscaling has been observed
in all of above mentioned  reactions, 
and one of the unifying features has been
the dominant role of the symmetry energy.
Because of this, isoscaling has been proposed as a
signal to be used in exploring the symmetry energy.

In this paper, we  examine  the  basis for isoscaling in the 
yields of fragments in binary fission.
That process differs from other reactions in that it
involves low energies and is strongly constrained by
mass and charge conservation.  The former eliminates
pre-equilibrium effects, and the later well characterizes the
portion of the system which remains after the observed
fragment leaves the total system, i.e., the complementary
fragment. 
Furthermore in fission, as compared to other reactions, the observed
fragment can represent an appreciable portion of the the total system.

The question of relative isotopic yields 
is particularly
relevant to the task of finding efficient methods for
populating nuclear species far from the valley
of stability. Interest in that subject is  prompted by the goal of producing 
isotopes near the neutron drip
line\cite{ber,soul,benll}--a region which includes  nuclei important for the
r-process of nucleosynthese. 

We first propose a simple model for estimating
the isotopic yields and use this model to  study isoscaling.  
The model
suggests  a dominant role for  the 
nuclear symmetry energy, as was the case with the other
reactions studied. We demonstrate that this
model is consistent with fission data in the literature\cite{lbl}, 
and show how
it can provide for isoscaling when the ratio is taken
of the yields of isotopes from
the fission of different parent nuclei.

We begin by assuming that, in the fission process, an excited nuclear
system divides into two fragments, each characterized by proton and
neutron numbers.  These fragments may be excited and lose a few
additional neutrons by evaporation after fission. 
Our goal, however, is to predict the isotopic
distributions of the
primary partition. Thus, detailed comparison with data
may require  modification to account for 
evaporation which leads to the fragments which are actually  observed

We assume that the isotopic yields in the fission fragments are
governed
by the conditions at  scission. A detailed model\cite{wilk} 
following  this approach
has previously been examined in the literature. In this work, however, 
we are concerned with a  special feature
of the processes, namely, how the neutrons of the fissioning system
are  partitioned between the two fragments, given the partition of
the protons.
The partition of neutrons provides the 
isotopic distribution of each element. 
We assume that, at scission, the system is in equilibrium
so that the probability for a given partition is given by the 
Boltzmann factor ${\rm exp}(-E_{part}/T)$. The energy $E_{part}$
consists of terms which reflect the binding energy of the individual
nuclei of the scissioning pair, and also terms related to the interaction
between the members of the pair.  
Whereas the interaction terms (including the long ranged
Coulomb interaction ) are important for the charge partition, 
we assume that they  only play a small role in determining the
partition of the neutrons. Thus, for finding the relative isotopic yields
for a given element, we ignore the
interaction terms and assume that the neutron distribution is
provided by 
the  binding energies, $BE_i$, of the two fragments
divided by a temperature $T$.
Thus, for a given charge partition:
\begin{equation}
 Y(z_1,n_1:z_2,n_2)\propto {\rm exp}((BE_1(z_1,n_1)+BE_2(z_2,n_2))/T),
\end{equation}
where $z_1+z_2=Z$ and $n_1+n_2=N$, and $Z$ and $N$ are charge and
neutron number of the
fissioning system. 
For the fission of heavy elements both
fission fragments are neutron rich.
Thus the respective variations in the
associated neutron number for the two systems will influence
the binding energies in opposite directions, i.e.,  more neutrons for
fragment 1 will reduce its binding energy, and, correspondingly,
fewer neutrons for the complementary fragment, 2,
will raise its binding energy. 
Under the assumption of Eq.2,
the maximum in the neutron distributions
for a given proton partition
will occur when the
\begin{equation}
\partial (BE_1(z_1,n_1)+BE_2(z_2,N-n_1))/\partial n_1=\partial BE_1(z_1,n_1)/
\partial n_1-\partial BE_2(z_2,n_2)/\partial n_2=0.
\end{equation}
 
We will first study the case when the binding energies 
are modeled by the terms
in a global liquid-drop model--volume, surface, coulomb, and symmetry--
with conventional coefficients\cite{bm}. Following this,  
we will a 
examine the effects produced by  the
addition of shell corrections.
Using the liquid-drop terms alone we find that
the overwhelmingly dominant contribution to the changes in the two binding
energies is provided by the respective symmetry energies. With this
term alone the requirement that the total binding be a
maximum leads to the condition  that  $(z_1/a_1)=(z_2/a_2)=(Z/A)$.
This follows directly from Eq.3 with the specific dependence 
of the symmetry term on neutron
number given by,
$((n-z)^2/a)$. The maximum of the isotope yields will be far from the 
the valley of stability
since  the  fissioning systems generally are more neutron rich than
either of the most stable isotopes of the resulting two elements. 
If, in addition to the symmetry term, the volume, surface, and coulomb 
terms of the liquid-drop formula
are included in the respective binding energies,
the predicted positions of the 
peaks of the isotope distributions are found to shift by less than  one
unit of mass (neutron).  
We will show below that
the maxima  can be further
shifted by the addition of the shell contributions to  the
binding energy. The observed peaks corrected for secondary evaporation,
do indeed show\cite{wahl} that the maximum is extremely close to the value
arising from the symmetry energy alone. 
This confirms the dominance of that term.  We note that in all observed cases
the maximum in the isotope distribution, as expected,  
is well removed from the valley of stability.

In addition to estimating  the peak in the isotope  distribution, we
also can estimate the widths for the distributions with Eq.2.
A value for this quantity  can be found by expanding the
total binding energy, given by the respective liquid-drop estimates, 
about the peak values. One finds here  that 
the symmetry energy term is again dominant. In fact while the other
terms in the liquid-drop contributions move the peak position slightly,
they have no noticeable effect on the width. We thus obtain 
a good approximation for the Gaussian width $\sigma$
 from the
symmetry terms alone,

\begin{equation}
\sigma^{-2}=8(C_{sym}/T)(Z/A)^3(Z/(z_1z_2)).
\end{equation}
Here $C_{sym}$ is the coefficient of the symmetry term in the binding
energy (generally on the order of 23 MeV \cite{bm}), and $T$ is the equilibrium
temperature introduced above.  

In Fig. 1a and 1b. we compare the observed 
distribution for the independent yields 
of complementary fragments
of $z_1=30$ and $z_2=62$ obtained from the asymmetry fission of
$^{234}U$ following the absorption of 14 $MeV$ neutrons
on $^{233}U$\cite{lbl}.  
The lines in the figure indicate the predictions using Eq.4
with three values of temperature, T=1.7,1.8, and 1.9 MeV.
In each of the two distributions  the peak positions of the
calculations have
been shifted 
(.75 mass units for the
lighter element and 2.0 mass units for the heavier element). These shifts
probably  reflect the effects of evaporation.
The slight asymmetry in the observed
distributions, where  diminished values are found  for the most neutron-rich
isotopes, is also consistent with the greater tendency for the
very neutron-rich primary isotopes to lose more neutrons by evaporation. 
The values for the fitting temperature are consistent with excitation energies
of 35-40 MeV\cite{bw}. Ground state
Q-values and
TKE systematics\cite{vv} would provide about 20 MeV. 
Additional energy is introduced  by the neutrons to provide the excitation
indicated.

We next take up the phenomenon of isoscaling.  
and begin
the  study  within the context 
of the liquid-drop model.
We   
predict that isoscaling will occur and 
derive expressions for the  values for the parameter $\alpha$ 
in the exponential
expression of Eq. 1. In the study of isoscaling in other types
of reactions simple arguments based on the liquid-drop model were
sufficient to obtain a good understanding of the isoscaling signal.
That signal was even used to learn about the symmetry part of the
the energy\cite{ts1}.

As a concrete illustration  we 
compare the isotope yields for two
fission 
processes, one for $^{239}U$ and the other for $^{234}U$,
characterized as heavy ($h$) and light ($\ell$).
In
calculating the isoscaling ratios $R_{h\ell}(n_1,z_1)$ for the isotopes of 
neutron number $n_1$, and proton number $z_1$ , the factors in the expression 
for yield in Eq. 2  which
involves
$BE_1(z_1,n_1)$
cancel, and the properties of the complementary
fragments, which are  different for the two fissioning systems,
determine $R_{h\ell}$,
\begin{equation} R_{h\ell}(z_1,n_1)\propto {\rm
exp}((BE_{2_{h}}(z_{2_{h}},n_{2_{h}})-BE_{2_{\ell}}(z_{2_{\ell}},
n_{2_{\ell}}))/T).
\end{equation}
Here $z_{2_{h}}=Z_{h}-z_1$, $z_{2_{\ell}}=Z_{\ell}-z_1$,
$n_{2_{h}}
=N_h-n_1$, and 
$n_{2_{\ell}}=N_{\ell}-n_1$. 
To reiterate, the  individual isotope distributions depend on 
both of the binding energies, but, in the ratio of the yields,
only the binding energy of the fragments which are complementary to
the one whose yield is considered are important.

To determine
the isoscaling parameter, $\alpha$, we consider the change in $R_{h\ell}$
with the change in $n_1$. This is directly related to the difference
in the separation  energies of the two complementary fragments.
The parameter $\alpha$ is thus well approximated from  the
symmetry energy term by:
\begin{equation}
\alpha=4(C_{sym}/T)[((Z_{\ell}-z)/(A_{\ell}-a))^2-((Z_{h}-z)/(A_{h}-a))^2].
\end{equation}
Here $A_{\ell}$ and $Z_{\ell}$ are the respective mass and charge of the 
lighter fissioning system 
and
$A_h$ and $Z_h$ the mass and charge of the heavier system, while
$a$ and $z$ are the mass and charge of
the specific isotope whose yields are compared in the ratio $R_{h\ell}$.
For the specific case of fission from  two  isotopes of a given element (Z)
with masses given respectively by $A_h$ (heavy) 
and $A_{\ell}$ (light), 
the prediction for $\alpha$ is
well represented  by the approximate expression
\begin{equation}
\alpha(z)=8(C_{sym}/T)(A_h-A_{\ell})(2Z/(A_h+A_{\ell})^3)/(Z-z).
\end{equation}

Notice that the predicted value of $\alpha$ increases with $z$. This
type of variation was not noticed in other types of reactions since
the charges of the observed fragments did not cover as large a portion of the
the entire system as they do in the fission process.
Using as an example the fission of $^{239}U$ and
$^{234}U$, we show in Fig. 2 the  values of $\alpha$ obtained from Eq.6.
The plot clearly shows the z dependence.
The curves represent
three values of temperature,
1.7, 1.8, and 1.9 MeV. 
These are the same values
shown in Fig. 1 for predictions of the isotope distributions. 
It is known that the effective symmetry 
coefficient contains surface
effects\cite{bm}and thus depends on the mass of the 
nucleus in consideration.
Under this circumstance it is the  $C_{sym}(a)$ for the complementary
fragment which determines the value of $\alpha$ in Eqs.6 and 7. 
In figure 2. we 
used the values of $C_{sym}(a)$ provided by 
parameters from a recent study\cite{desou}.

The discussion up to this point has been based on the
the use of the liquid-drop model for the binding
energies  of each of the binary fragments.
This procedure  provided the predictions for the 
isotope distributions and also for the isoscaling
parameter $\alpha$.  
For the case of fission, however, 
there may be additional features.
These arise from the
fact that the energy is relatively low and from the constrains of
mass and charge conservation which proscribe the features of the  system 
complementary to the observed fragments.
We discuss two of these effects next.

The exact configuration of the fragments and their deformation
at scission is not known.  We assume, however,  that
the contributions to the liquid-drop energies
will be little affected by these considerations. 
However, additional
detailed structural features, such as shell effects, can also 
affect the binding energies,
but they may be more influenced by the specific nature of the scission
configuration.
It is, nonetheless, instructive to explore the possible influence
of shell corrections, even if the exact form is unknown.
For this purpose we have
examined the differences between the values of the binding energies 
tabulated in the literature\cite{masstab} for free nuclei 
and the predictions of the simple liquid-drop. This gives an
indication of the role of such effects.
The differences in binding energies  include pairing corrections
as well as contributions to arising from the closing of nuclear shells.
We nonetheless refer to these differences
here as ``shell corrections'', and note that actual effects at scission may
differ from those for free nuclei.

For the fission fragments of interest,
one finds, as expected, that the differences are greatest 
in the vicinity of  magic  numbers for the neutrons,
50 and 82.
One of the consequences of the shell
contributions to the binding energies
is a  shift in the location of the
peaks of the isotope distributions from the values predicted by liquid-drop 
considerations.

We consider the situation for the fission of $^{234}U$, as an
example. In Fig.3a,
the size of the shifts in the peaks in the isotope distributions
relative to the values
predicted by the symmetry energy alone are plotted.
We have calculated the peaks of the isotope distributions
assuming that 
yields are governed by ${\rm exp}(BE_1+BE_2)/T$, and
we have taken the values of $BE$ for each
nuclide from a standard 
mass tables\cite{masstab}.
This procedure
can only be performed for a limited number of isotopes
because the  tables are incomplete.
The values  for the shifts show opposite signs for the 
heavy and
light members of the pair of fragments as required by particle conservation.
Only results for even $z$ are shown  
to suppress additional fluctuations due to
pairing. One finds that the largest mass shifts are 
approximately 3 units, and these occur for the pair with charges equal to
50 and 42.
This case occurs  when the charge for the heavier fragment is 50
which  has, at the peak of the isotope distribution,  
a neutron number of 82 (a closed shell).
For other pairs of fragments,  the shift in the peak is smaller

The differences in binding energy (tabulated energies
minus liquid-drop energies)
have been evaluated for isotopes at the peaks
predicted by the tabulated energies. These differences
are plotted in Fig.3b. The largest difference also occurs
for the pair with charges equal to 50 and 42
where the change in energy is approximately 8 MeV.

The shell corrections
are found to modify the prediction for the 
width of the isotope distributions from values obtained
using the symmetry energy alone.
This modification is found to give a reduction 
on the order of 20\% in width for the fragments for the binary
pair with charges 50 and 42.

We next examine how the shell corrections effect 
the isoscaling signal.  With only the liquid-drop contributions, 
the dependence  of log$(R_{h\ell}(n,z))$ on $n$ for the yields
for a given z is
approximately linear (assuming the temperatures at scission are the same).
That is, for a given $z$ the
ratio is expected to follow exp$(-\alpha n)$, where $n$
runs over the neutron numbers of the different isotopes.
This is a necessary condition for isoscaling.  
Shell corrections can modify this behavior, however.   
In our model, the value of $R_{h\ell}$ is  determine by the
binding energy 
of the two fragments complementary to the one whose yield is involved in the
ratio. The value of $\alpha$ reflects the difference between the separation
energies for these two nuclei.
These respective separation energies are  influenced  by shell
effects. In the case of the fissioning of systems of
different neutron number, the
neutron numbers of the  complementary fragments will differ by the same
value as the difference in the total neutron numbers 
for the two fissioning systems. 

For the case of the yields from $^{234}U$ and $^{239}U$, for example, the 
neutron numbers for the complementary
fragments differ by 5 neutrons.   The shell closures for these
two nuclei will consequently be apparent in the yield
ratios for  values of $n$ separated
by 5 neutrons.
Between the shell closure values the binding
energy for one of the complementary systems will be rising while the
other is falling. This has as very strong effect of the n-dependence of
log$(R_{h\ell}(n,z))$.
In particular the curves
will deviate shapely from the linear form associated with the liquid-drop case.

We have examined this effect through an example involving  
ratios of yields from
the two Uranium isotopes.  The result is a sharp change in the slope
of the log($R_{h\ell}(n,z)$. 
The shell corrections
place this change in the region of neutron numbers running between
$n=60$ (where the complimentary fragment as $N=82$
for $^{234}U$) and  $n=65$ (where the complementary fragment has $N=82$ for
 $^{239}U$).  The  calculation of this behavior for log$(R_{h\ell})$
is shown in Fig.4  where the tabulated masses, rather than
the liquid-drop masses, have been inserted in Eq.5 for $R_{h\ell}$.
Because the mass tables are incomplete this procedure
can only be performed for the limited number of isotopes shown in the figure.
The dash line represents the behavior of log$(R_{h\ell})$ predicted with 
liquid-drop masses.  
When the shell corrections are included the extraction of an $\alpha$ is  
uncertain because of the changing slope.
Even outside of the region of the steep rise,
the values of $R_{h\ell}$
are still influenced by the differences
between the shell effects in the two systems. This feature can change
the smooth dependence of $\alpha$ on $z$ found with the liquid-drop masses.
One would anticipate that these deviations would
be greatest for those values of $z$ 
which involve neutron numbers in the vicinity of 60-65
where the shell effects are expected to be largest. In a recent
preprint Veselsky {\it et al.}\cite{ves} 
have presented  observations of some of these
features in the fission data base of ref.7. Their
interpretation of the effect is, however, very different
from what we present here.  

We briefly review the features we have found for the prediction of the
isoscaling parameter $\alpha$.
If only the liquid-drop energies are used for the fission
fragments one would expect to find a smooth linear dependence on $n$
for log$(R_{h\ell})$ and values of logarithmic slope
will vary approximately like $1/(Z-z)$.
If additional contributions to the energies are involved, 
such as those arising from shell effects,
the behavior of $R_{h\ell}$ can be radically
affected.  This affect makes the value of the isoscaling parameter uncertain
and this may account for some of the effects reported
by Veselsky\cite{ves}.
One can predict that this will occur in regions affected by
the large shell effects. Even for values of $n$ beyond that of the rapid
rise in $R_{h\ell}$, where the dependence returns to that
the liquid-drop values, the slopes and the apparent value
of the isoscaling parameter, $\alpha$, may deviate due to the 
remaining influence of 
the shell corrections. This can even cause the apparent values of $\alpha$
to decrease with increasing $z$, as appears to be the case in Fig.4. 
This would occur in a narrow region 
around $z=40$ for the fission of the two 
Uranium systems.  At values
$z$ distant from these, the shell effects fade and the general trend
in the $z$ dependence of $\alpha$ associated with the liquid-drop
energies is reestablished.

In summary, the studies in this work
suggest several properties for the isotopic yields and the isoscaling
signals. Following from  the assumption that the isotopic yields
in fission are governed by the total binding energy at scission,
we have
found that the contribution from symmetry energy  is primarily responsible
for the location of the peak value in the isotopic  distribution.
For each element this peak value  
is provided by the ratio (Z/A) of the fissioning
system.   This is well satisfied for observed yields in the
literature\cite{wahl}.  The widths of these distributions 
have been also been shown to be related to the
strength of the symmetry energy, a temperature, and simple 
factors depending on the
the proton and mass numbers of the fragments.  For the case of the
two complementary distributions (Z=30 and Z=62) arising from the asymmetric
fission of $^{234}U$ (Z=92), the agreement with the independent
yields  are consistent with a shift
in the peak position of one or two neutrons, and with a temperature of 
approximately 1.7-1.9 MeV. This agreement is achieved for
this pair of $z$ values under the assumption that the liquid-drop model
well represents the binding energies and that shell effects
are unimportant. A slight asymmetry in the tabulated
independent yields is consistent with 
increased secondary evaporation for the most neutron-rich isotopes.
The values
for the 
temperatures are  reasonable according to
the general energy balance.

The isoscaling behavior depends on features peculiar to the
fission process. The scission-energy model with liquid-drop energies
provides predictions for
values of the isoscaling parameter $\alpha$.
We found 
that $\alpha$  is expected to
increase with increasing proton number of the observed element.
This increase is apparent because of the large
range of elements observed in fission.
For the comparison of isotopes from the fission of $^{234}U$ and $^{239}U$, 
values of $\alpha$ would  range from about  0.40 to 1.0 
over the accessible values of $z$ for the values of  temperature
which
provide agreement with corresponding observed isotope 
distributions.
Because of the low energies and the strong constraint on the
system complementary to the observed fragments, shell effects
can also affect isoscaling in fission. 
The influence of shell effects especially near N=50 and N=82
modify the isotope distribution in peak position and in width.
Furthermore the shell effects can have a very strong  effect 
on $R_{h\ell}$.
These corrections can make uncertain  the determination of 
an isoscaling parameter, $\alpha$, and they  can affect its apparent 
values, causing them to deviate from the smooth behavior
associated with liquid-drop binding energies. 

This work was supported in part by grants from the US National Science
Foundation, PHY-0070161.



\newpage

\begin{figure}
\includegraphics[height=5in,angle=90]{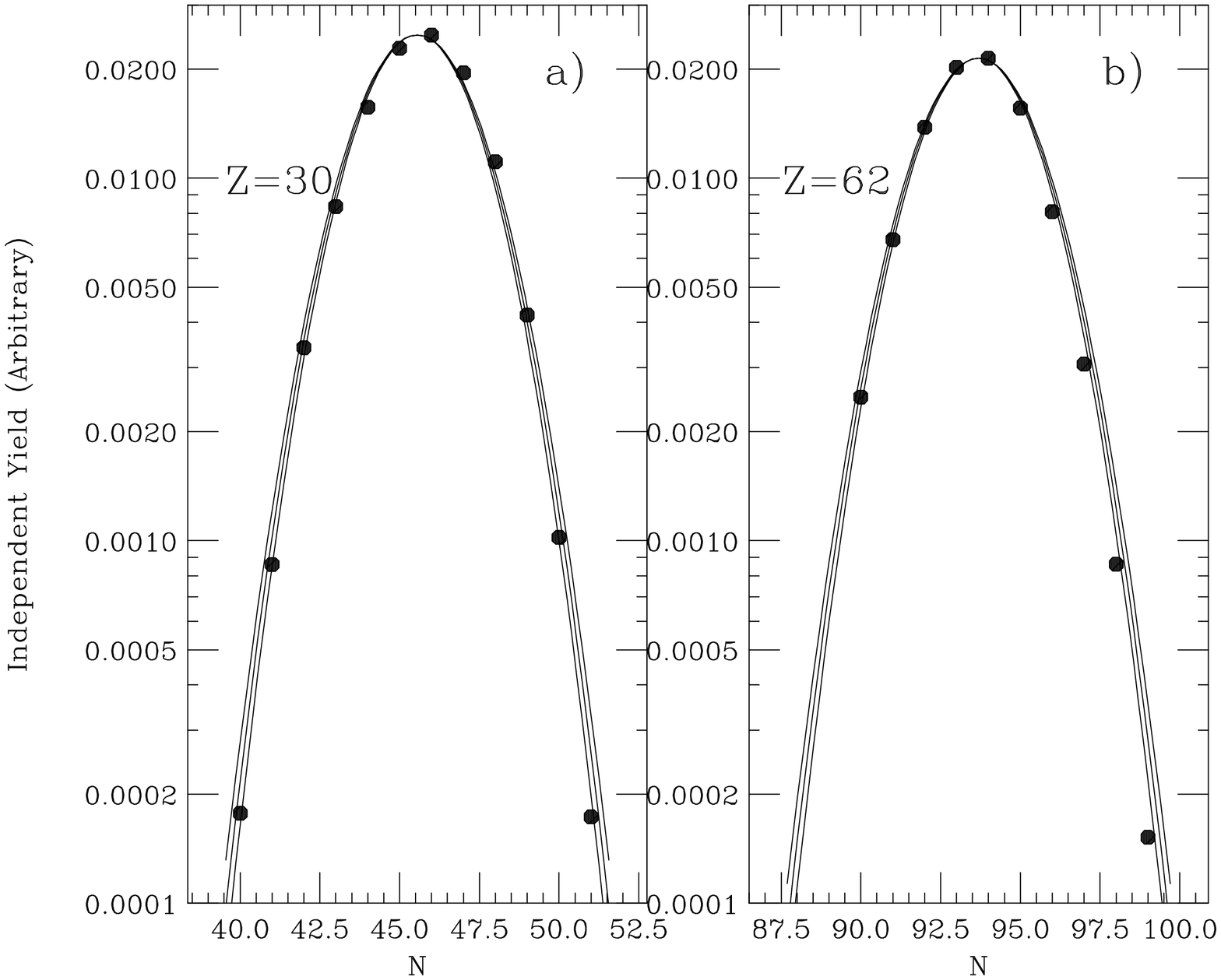}
\caption{Neutron distribution for two complementary fragments
from the fission of $^{234}U$. 
a) Z=30; b) Z=62. Points from Ref.\cite{lbl}; curves calculated from
Eq.4 with T=1.7,1.8,1.9, and peaks shifted from symmetry energy values
by  .75 MeV in a) and 2.0 MeV in b)}
\label{F1}
\end{figure}

\begin{figure}
\includegraphics[height=5in,angle=90]{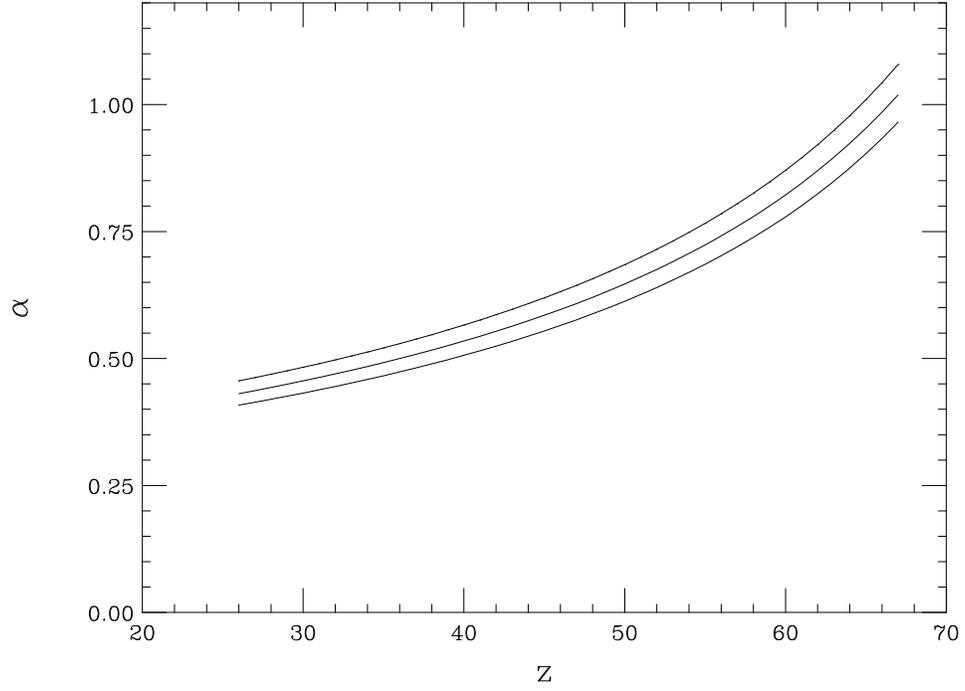}
\caption{ Calculated values of isoscaling parameter $\alpha$
as a function of the proton number of the fission fragments
from  $^{239}U$ relative to $^{234}U$
obtained from Eq.6 with T=1.7,1.8,1.9 MeV (top to bottom).}
\label{F2}
\end{figure}

\begin{figure}                        
\includegraphics[height=5in,angle=90]{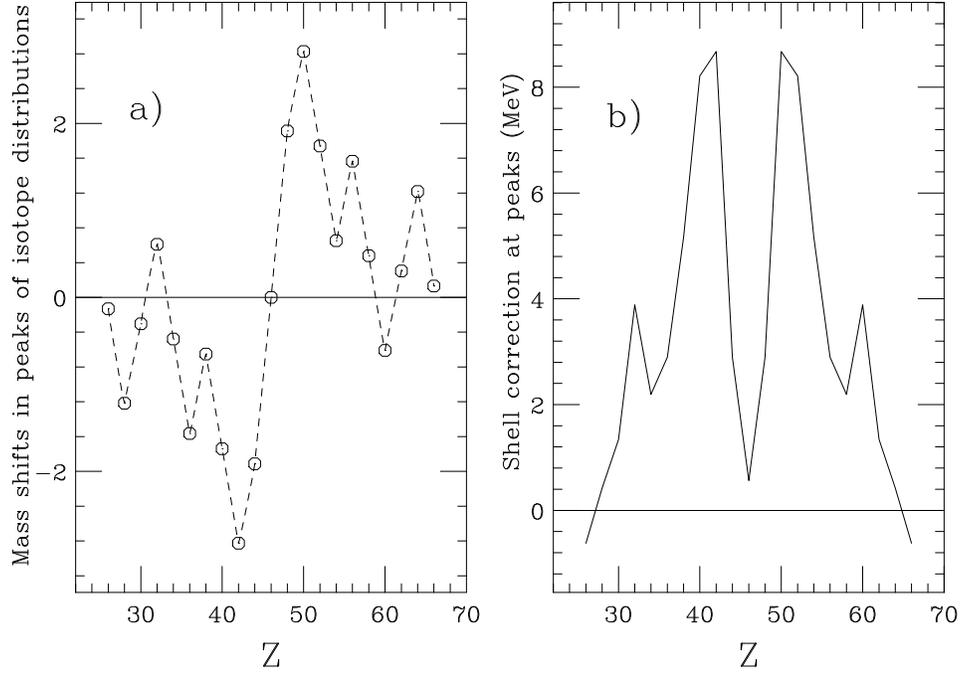}
\caption{a) Shifts in the mass number of the peaks in the isotope distributions
for fragments from the binary fission of $^{234}U$.
Open circles give the difference between peak position using tabulated mass 
values in Eq.2.
relative to those obtained with 
the symmetry energy of the liquid-drop model. b) The energy
differences between the values from
mass tables and values from the liquid-drop model 
for the isotopes at the peaks of
the distributions. Only fragments for even $z$ are indicated in both figures.}
\label{F3}
\end{figure}

\begin{figure}
\includegraphics[height=5in,angle=90]{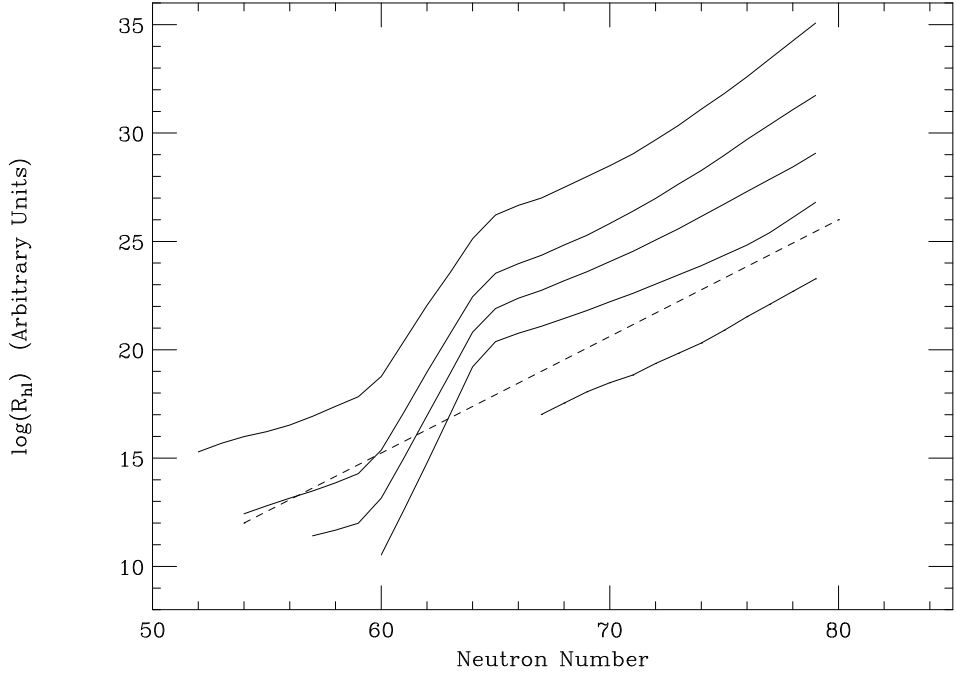}
\caption{ Calculated values of log$(R_{h\ell}(n,z))$
for $z=35,38,40,42,45$ (top to bottom) from the fission of $^{239}U$
and $^{234}U$. Binding energies
from mass tables are used in  Eq.2. The dashed curve indicates
the result for z=40 with liquid-drop masses.
The scale is in arbitrary units and neighboring isotopes are
averaged to suppress odd-even fluctuations.}
\label{F4}
\end{figure}

\end{document}